\documentclass[letterpaper]{article} 
\usepackage{aaai2026}  
\usepackage{times}  
\usepackage{helvet}  
\usepackage{courier}  
\usepackage[hyphens]{url}  
\usepackage{graphicx} 
\usepackage{natbib}  
\usepackage{caption} 
\usepackage{algorithm}
\usepackage{algorithmic}

\usepackage{amsmath}
\usepackage{graphicx} 
\usepackage{booktabs}
\usepackage{multirow}
\usepackage{amssymb}
\usepackage{array} 
\usepackage{makecell}
\usepackage{newfloat}
\usepackage{listings}
\DeclareCaptionStyle{ruled}{labelfont=normalfont,labelsep=colon,strut=off} 
\floatstyle{ruled}
\newfloat{listing}{tb}{lst}{}
\floatname{listing}{Listing}
\title{FourierPET: Deep Fourier-based Unrolled Network for Low-count PET Reconstruction}
\author{
    Zheng Zhang\textsuperscript{\rm 1},
    Hao Tang\textsuperscript{\rm 1}\thanks{Corresponding author.},
    Yingying Hu\textsuperscript{\rm 2},
    Zhanli Hu\textsuperscript{\rm 3},
    Jing Qin\textsuperscript{\rm 1}
}
\affiliations{
    \textsuperscript{\rm 1}Centre for Smart Health, The Hong Kong Polytechnic University\\
    \textsuperscript{\rm 2}The Department of Nuclear Medicine, Sun Yat-sen University Cancer Center\\
    \textsuperscript{\rm 3}Research Center for Medical AI, Shenzhen Institute of Advanced Technology, Chinese Academy of Sciences\\
    zheng1.zhang@connect.polyu.hk, \{howard-hao.tang, harry.qin\}@polyu.edu.hk, huyy@sysucc.org.cn, zl.hu@siat.ac.cn
}

\usepackage{bibentry}

\begin{document}

\maketitle

\begin{abstract}
Low-count positron emission tomography (PET) reconstruction is a challenging inverse problem due to severe degradations arising from Poisson noise, photon scarcity, and attenuation correction errors. 
Existing deep learning methods typically address these in the spatial domain with an undifferentiated optimization objective, making it difficult to disentangle overlapping artifacts and limiting correction effectiveness.
In this work, we perform a Fourier-domain analysis and reveal that these degradations are spectrally separable: Poisson noise and photon scarcity cause high-frequency phase perturbations, while attenuation errors suppress low-frequency amplitude components. Leveraging this insight, we propose \textit{FourierPET}, a Fourier-based unrolled reconstruction framework grounded in the Alternating Direction Method of Multipliers. It consists of three tailored modules: a \textit{spectral consistency module} that enforces global frequency alignment to maintain data fidelity, an \textit{amplitude–phase correction module} that decouples and compensates for high-frequency phase distortions and low-frequency amplitude suppression, and a \textit{dual adjustment module} that accelerates convergence during iterative reconstruction. Extensive experiments demonstrate that \textit{FourierPET} achieves state-of-the-art performance with significantly fewer parameters, while offering enhanced interpretability through frequency-aware correction. Our code is available at: \url{https://github.com/xiaochaorouz/FourierPET}.
\end{abstract}


\section{Introduction}
\begin{figure*}[t]                  
  \centering
  \includegraphics[width=\textwidth]{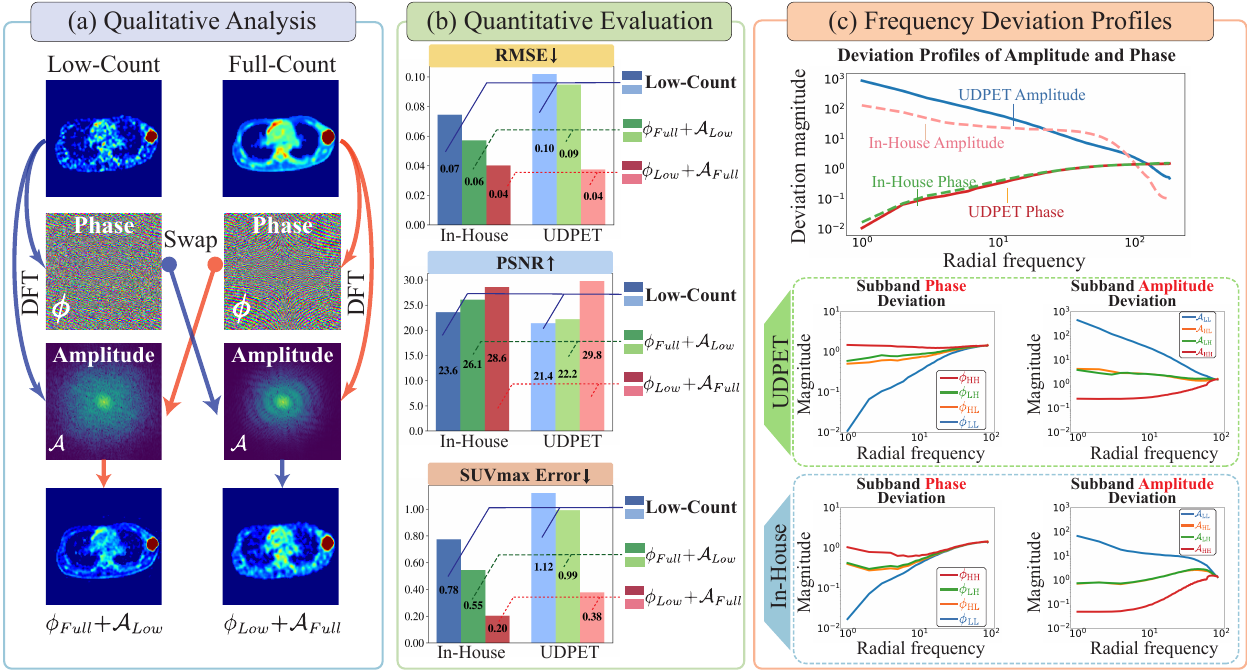}\vspace{2mm}
    \caption{\textbf{Motivation.} \textbf{(a)} Qualitative and \textbf{(b)} quantitative analyses demonstrate that low-count degradations exhibit separable spectral patterns: Poisson noise and photon starvation mainly perturb the \emph{phase}, degrading structural fidelity, while attenuation-correction (AC) errors suppress low-frequency \emph{amplitude}, inducing global intensity bias. \textbf{(c)} Frequency-deviation profiles reveal that phase errors concentrate in high frequencies, whereas amplitude distortions dominate low-frequency bands.}
  \label{fig:total_experiment}
\end{figure*}

Positron emission tomography (PET) is a molecular imaging modality widely used in oncology and neurology to visualize abnormal metabolic activity. It reconstructs radiotracer distributions from coincident photon measurements along lines of response (LORs). To reduce radiation dose and scan time, clinical protocols often operate in \emph{low-count} regimes, which degrade reconstruction quality through three main factors: (i) Poisson noise lowers the signal-to-noise ratio (SNR); (ii) photon starvation diminishes fine structural detail~\cite{yan2016method}; and (iii) attenuation correction (AC) errors introduce systematic intensity bias~\cite{wang2020machine,chen2017attenuation}. Although these effects have distinct physical causes, they are intertwined in the image domain, making targeted compensation difficult.

Existing PET reconstruction methods fall into three categories: (i) iterative algorithms with physics-based priors~\cite{hudson1994accelerated,shepp2007maximum,hutchcroft2016anatomically}; (ii) end-to-end networks mapping sinograms to images~\cite{zhang2023deep,xie2025prompt,wang2020fbp,kaviani2023image,cui2024mcad,hu2022transem}; and (iii) post-hoc refinements~\cite{han2023contrastive,tang2024hf,xue2025deep}. Despite their success, most methods address degradation effects in an undifferentiated manner, without exploiting potential separability in representational space. Spectral modeling, which has facilitated such separation in other modalities~\cite{haller2021susceptibility,zhou2023fourmer,zhou2024seeing}, has seen limited exploration in PET. This motivates a central question: \textit{Can we design frequency-aware models that selectively isolate and correct degradations rooted in distinct physical processes?}

\paragraph{Key observation.}
Our analyses reveal that, despite spatial entanglement, low-count PET degradations manifest \emph{separable spectral patterns}. Specifically, Poisson noise and photon scarcity induce high-frequency phase perturbations that degrade structural sharpness, while AC bias introduces smooth multiplicative fields that suppress low-frequency amplitude. We validate this with clinical data (Fig.~\ref{fig:total_experiment}): (1) high-frequency phase variance intensifies with reduced counts due to stochastic noise; (2) low-frequency amplitude is systematically attenuated by AC bias. These findings suggest that targeting amplitude and phase components \emph{separately} may offer a more principled correction strategy than unified image-domain penalties.

\paragraph{Our idea.}
We make the spectral factorization \emph{actionable} in a model-based objective. Let \(x\) denote the reconstructed PET image, \(\mathrm{A}\) the PET system matrix, and \(y\) the measured sinograms. We formulate
\begin{equation}
\min_{x}\;
\underbrace{\mathcal{L}(\mathrm{A}x, y)}_{\scriptstyle \text{data fidelity}} +
\lambda_a\, \underbrace{\mathcal{R}_{\text{amp}}\!\big(\lvert\mathcal{F}(x)\rvert\big)}_{\scriptstyle \text{\shortstack{LF amplitude\\correction}}} +
\lambda_p\, \underbrace{\mathcal{R}_{\text{phase}}\!\big(\angle\mathcal{F}(x)\big)}_{\scriptstyle \text{\shortstack{HF phase\\stabilization}}},
\end{equation}
where \(\mathcal{F(\cdot)}\) is the Fourier transform, \(|\cdot|\) and \(\angle(\cdot)\) extract amplitude and phase, and \(\mathcal{R}_{\text{amp}},\mathcal{R}_{\text{phase}}\) are spectrally targeted priors tailored to AC bias and low-count noise, respectively.

\begin{figure*}[t]                  
  \centering
  \includegraphics[width=\textwidth]{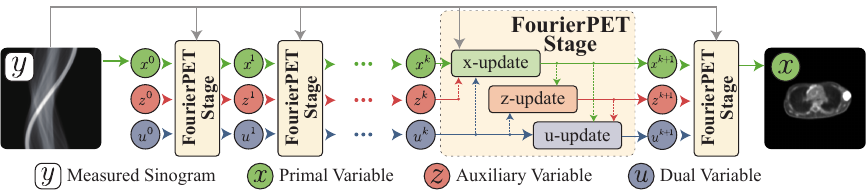}
    \caption{\textbf{Overview of the proposed \textit{FourierPET} architecture.} Given a measured sinogram $y$, \textit{FourierPET} performs $K$ unrolled ADMM iterations to iteratively refine the primal variable $x$, auxiliary variable $z$, and dual variable $u$. Each iteration comprises three steps: (1) the $x$-update enforces measurement consistency and spectral alignment; (2) the $z$-update applies frequency-aware regularization to mitigate degradation; (3) the $u$-update promotes convergence by reconciling $x$ and $z$. The final reconstruction is obtained after $K$ iterations.}
  \label{fig:flowchart}
\end{figure*}

\paragraph{\textit{FourierPET}.}
To solve this objective, we derive an Alternating Direction Method of Multipliers (ADMM)-based variable-splitting scheme and unroll it into a learnable architecture (see Fig.~\ref{fig:flowchart}): (1) $x$-update (\textbf{Spectral Consistency Module (SCM)}) that enforces data fidelity via the system matrix and promotes global frequency alignment through state-space Fourier neural operators (SSFNO); (2) $z$-update (\textbf{Amplitude–Phase Correction Module (APCM)}) that explicitly restores low-frequency amplitude suppressed by AC errors and stabilizes high-frequency phase degraded by Poisson noise and photon scarcity; and (3) $u$-update (\textbf{Dual-Adjustment Module (DAM)}) that dynamically coordinates $x$ and $z$ variables to accelerate and stabilize convergence. This design preserves the interpretability of model-based optimization while injecting frequency-selective, physically motivated corrections.

\paragraph{Contributions.}
(i) We present a frequency-domain perspective linking low-count degradations to distinct \emph{amplitude/phase} patterns, validated across multiple PET datasets.  
(ii) We propose \textit{FourierPET}, an ADMM-unrolled framework that integrates spectral data fidelity with directional priors for interpretable, frequency-aware correction.  
(iii) We design \textbf{SCM} to enforce reconstruction fidelity in both spatial and frequency domains; \textbf{APCM} to rectify \emph{low-frequency amplitude} suppression from AC bias and \emph{high-frequency phase} drift from low-count conditions; and \textbf{DAM} to ensure stable and efficient convergence.  
(iv) Extensive experiments show improved fidelity, accuracy, and robustness, with ablations confirming the impact of spectral targeting.

\section{Preliminary Study}
\noindent\textbf{Motivation.} 
Low-count PET is degraded by three intertwined factors: (i) Poisson noise from reduced photon statistics, (ii) photon starvation due to low-dose or shortened acquisition, and (iii) systematic AC bias induced by anatomical mismatch. These effects are spatially entangled and strain conventional reconstruction. We posit that, although entangled in the image domain, these degradations manifest \emph{separable spectral signatures} when the image is decomposed into amplitude and phase.
Intuitively, count fluctuations predominantly perturb edge geometry and fine structures (captured by phase), while AC-related gain shifts act as varying intensity modulations (captured by low-frequency amplitude). Guided by this hypothesis, we perform a three-stage spectral analysis (Fig.~\ref{fig:total_experiment}) to reveal actionable decomposition patterns.

\noindent\textbf{Amplitude–Phase Swap Analysis.}  
We begin by probing whether degradations can be spectrally disentangled. Specifically, we reconstruct hybrid PET images by interchanging amplitude and phase components between full-count and low-count scans using Discrete Fourier Transform (DFT)-based decomposition: 
\[
I_{\text{Hybrid}} = \mathcal{F}^{-1}(\mathcal{A}_{A} \cdot e^{j\phi_{B}}),
\]
where $\mathcal{A}_{A}$ and $\phi_{B}$ denote the amplitude and phase spectra from scans $A$ and $B$, respectively. 
Two hybrid variants are synthesized: (1) $\mathcal{A}_{Full} + \phi_{Low}$, and (2) $\mathcal{A}_{Low} + \phi_{Full}$. As shown in Fig.~\ref{fig:total_experiment}(a), phase substitution significantly reduces blurring and noise in metabolic regions but fails to restore contrast. Conversely, amplitude substitution recovers global intensity while leaving structural fidelity impaired. These observations suggest functional separation: \textit{photon statistics chiefly affect phase, whereas AC-related effects primarily alter amplitude}.

\noindent\textbf{Quantitative Evaluation.}  
To validate these qualitative insights, we conduct quantitative evaluations across two datasets: the UDPET dataset (206 subjects) and an in-house cohort (60 subjects). Four configurations are compared: (i) low-count baseline, (ii) phase-corrected ($\phi_{Full} + \mathcal{A}_{Low}$), (iii) amplitude-corrected ($\phi_{Low} + \mathcal{A}_{Full}$), and (iv) full-count reference. Metrics include PSNR, RMSE, and SUV$_\text{max}$ (surrogate for lesion detectability).
Fig.~\ref{fig:total_experiment}(b) shows that \textit{both correction strategies yield substantial gains over the low-count baseline, demonstrating complementary roles}.

\noindent\textbf{Frequency Deviation Profiling.}
 To further explore degradation localization in frequency space, we analyze spectral deviations of amplitude and phase over radial frequency bands. While DFT offers global analysis, Discrete Wavelet Transform (DWT) enables decomposition into four directional frequency subbands: high--high ($\mathrm{HH}$), high--low ($\mathrm{HL}$), low--high ($\mathrm{LH}$), and low--low ($\mathrm{LL}$). Our findings (Fig.~\ref{fig:total_experiment}(c)) reveal two consistent patterns: (1) phase variance concentrates in high-frequency $\mathrm{HH}$, indicating structure-related perturbations; (2) amplitude deviations dominate the low-frequency $\mathrm{LL}$ band, consistent with global gain shifts attributable to AC bias. These profiles provide empirical grounding for frequency-specific priors in downstream correction.

\noindent\textbf{Summary.}
Though low-count degradations appear spatially entangled, frequency analysis reveals two orthogonal failure modes: (1) \textbf{high-frequency phase disruptions} caused by photon scarcity and Poisson noise, and (2) \textbf{low-frequency amplitude suppression} induced by AC bias. This decomposition is both diagnostic and prescriptive: correcting each component along its spectral axis enables precise, interpretable improvements. We leverage this insight to design \emph{\textit{FourierPET}}, which explicitly regularizes high-frequency phase while correcting low-frequency amplitude bias.

\section{Methodology}
\paragraph{Problem Formulation.}
In low-count PET imaging, the objective is to reconstruct the underlying radiotracer distribution $x$ from sparse and noisy sinogram measurements $y$, which are modeled as $y = \mathbf{A}x + n$, where $n$ approximates combined measurement corruption from Poisson noise and electronic perturbations. This gives rise to the following inverse problem with regularization:
\begin{equation}
\label{eq:objective}
x^\ast = \arg\min_{x}\; \frac{1}{2}\|y - \mathbf{A}x\|_2^2 + \lambda\,g(x),
\end{equation}
where $g(x)$ imposes prior constraints to compensate for the ill-posedness of the reconstruction, and $\lambda > 0$ controls the trade-off between data fidelity and prior strength. 

\paragraph{Optimization via ADMM.} 
However, when $g(x)$ is non-differentiable or computationally complex, direct optimization of Eq.~\eqref{eq:objective} can be challenging. To address this, we introduce an auxiliary variable $z$, such that $x = z$, and solve the constrained optimization using the Alternating Direction Method of Multipliers (ADMM)~\cite{boyd2011distributed}. The augmented Lagrangian is:
\begin{equation}
\mathcal{L}_\rho(x, z, u) = \frac{1}{2} \|y - \mathbf{A}x\|_2^2 + g(z) + \frac{\rho}{2} \|x - z + u\|_2^2 - \frac{\rho}{2} \|u\|_2^2,
\end{equation}
where $u$ is the scaled dual variable and $\rho > 0$ is the penalty parameter. The ADMM iterations proceed as follows:
\begin{subequations}
\label{eq:subs}
\begin{align}
x^{k+1} &= \arg\min_x\; \frac{1}{2} \|y - \mathbf{A} x\|_2^2 + \frac{\rho}{2} \|x - z^k + u^k\|_2^2, \label{eq:sub1}\\
z^{k+1} &= \arg\min_z\; g(z) + \frac{\rho}{2} \|z - (x^{k+1} + u^k)\|_2^2,\label{eq:sub2}\\
u^{k+1} &= u^k + (x^{k+1} - z^{k+1}).\label{eq:sub3}
\end{align}
\end{subequations}

\paragraph{Deep Unrolling with \textbf{\textit{FourierPET}}.}
To combine the interpretability of iterative optimization with the representational power of deep learning, we propose \textbf{\textit{FourierPET}}, a learnable reconstruction network derived by unrolling $K$ iterations of ADMM into a feed-forward architecture. Each stage emulates one iteration of Eq.~\eqref{eq:subs}, preserving the modularity of ADMM while allowing neural components to be inserted into specific subproblems. 
Specifically:
\begin{itemize}
    \item \textbf{$x$-update}~(Eq.~\eqref{eq:sub1}): Performs a reconstruction step that ensures fidelity to the measured sinogram. We further enhance this step with global spectral refinement to eliminate measurement-inconsistent components.
    \item \textbf{$z$-update}~(Eq.~\eqref{eq:sub2}): Acts as a prior-guided regularization step. We introduce a domain-specific regularizer based on \emph{Frequency Deviation Profiling}, which compensates for low-frequency amplitude attenuation (to enhance contrast) and corrects high-frequency phase deviation (to suppress noise).
    \item \textbf{$u$-update}~(Eq.~\eqref{eq:sub3}): Coordinates between $x$ and $z$ through a learnable dual update, promoting stable convergence in the unrolled architecture.
\end{itemize}
An overview of our \textit{FourierPET} is illustrated in Fig.~\ref{fig:flowchart}. Each stage of the network explicitly mirrors one ADMM iteration, enabling structured updates that jointly enforce measurement consistency, spectral correction, and optimization convergence. This principled design leads to robust and high-fidelity reconstructions under low-count conditions.
\begin{figure}[t]      
  \centering
  \includegraphics[width=\columnwidth]{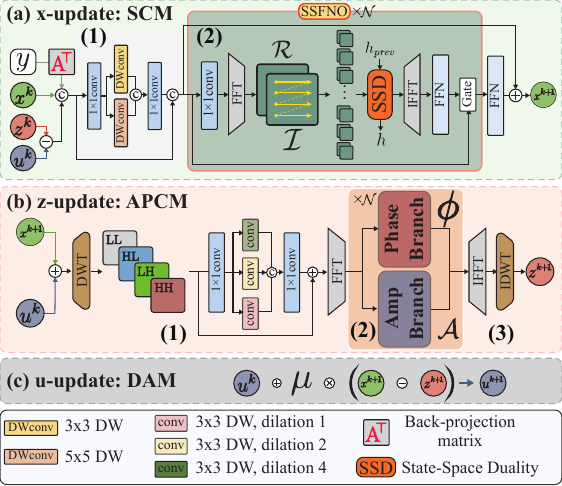}
  \caption{\textbf{Structure of a single \textit{FourierPET} iteration}, with three sequential components: \textbf{(a)} the $x$-update via SCM, \textbf{(b)} the $z$-update via APCM, and \textbf{(c)} the $u$-update via DAM.}
  \label{fig:x_z_u}
\end{figure}
\subsection{$x$-update: Spectral Consistency Module}
The $x$‑update step in Eq.~\eqref{eq:sub1} involves solving the following normal equation:
\begin{equation}
   x^{k+1}
= \bigl(\mathbf{A}^\top \mathbf{A} + \rho I\bigr)^{-1} \left( \mathbf{A}^\top y + \rho \left( z^k - u^k \right) \right),
\end{equation}
which is computationally expensive in large-scale PET reconstruction due to the matrix inversion. Although iterative solvers can provide numerical approximations, they often neglect critical structural and spectral priors—particularly problematic in low-count PET settings where noise severely degrades signal fidelity.

To address these limitations, we introduce the \textbf{Spectral Consistency Module (SCM)} as a learnable surrogate for the inverse operator $(\mathbf{A}^\top \mathbf{A} + \rho I)^{-1}$. SCM integrates domain knowledge through back-projection matrix $\mathbf{A}^\top$, ensuring consistency with measured sinogram $y$, and simultaneously learns to incorporate spatial and spectral priors essential for accurate reconstruction.

As shown in Fig.~\ref{fig:x_z_u}(a), SCM consists of two cascaded  components:
\textbf{(1) Spatial Module via DWConvs:} We first apply parallel depthwise-separable convolutions with kernel sizes $3\times3$ and $5\times5$ to the initialized features. These layers extract local metabolic structures across multiple scales, enhancing denoising capacity and robustness to low SNR.
\textbf{(2) Spectral Module via SSFNO:} The enriched spatial features are then fed into a stack of $\mathcal{N}$ \textbf{State‑Space Fourier Neural Operator (SSFNO)} blocks, which model global dependencies in the frequency domain. Specifically, we perform a Fast Fourier Transform (FFT) to obtain real and imaginary components $\mathcal{R}, \mathcal{I}$, which are flattened as $R' = \mathrm{Flatten}(\mathcal{R})$ and $I' = \mathrm{Flatten}(\mathcal{I})$, and subsequently processed by a State-Space Duality (SSD) module~\cite{lee2025efficientvim}:
\begin{equation}
\hat{R},\; \hat{I},\; h = \mathrm{SSD}\left( R',\; I',\; h_{\text{prev}} \right),
\end{equation}
where $h$ denotes a hidden recurrent state passed across stages for information flow and cross-stage consistency.

\paragraph{Remark.} The term ``Spectral Consistency'' reflects SCM’s ability to maintain coherence in the global frequency domain via SSFNO, while maintaining measurement consistency with the sinogram by incorporating $\mathbf{A}^\top$ as a fixed physical constraint at each iteration. This hybrid design enables SCM to approximate the inverse operator in Eq.~\eqref{eq:sub1} in a data-driven yet physically constrained manner, promoting both convergence stability and high-fidelity reconstruction.

\begin{figure}[t]      
  \centering
  \includegraphics[width=\columnwidth]{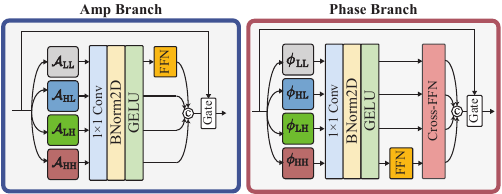}
  \caption{\textbf{APCM Core modules.} The Amp Branch (left) restores suppressed low-frequency components, while the Phase Branch (right) corrects high-frequency drifts. Together, these submodules provide targeted, frequency-aware compensation for degradations in low-count PET.}
  \label{fig:PA}
\end{figure}

\begin{table*}[t]
  \centering
  \begingroup
  \renewcommand{\arraystretch}{0.95}
  \setlength{\extrarowheight}{0pt}
  \footnotesize
  \begin{tabular}{l|ccc|ccc|ccc|c}
    \toprule
    \textbf{Method}
      & \multicolumn{3}{c|}{\textbf{BrainWeb (20\% Count)}}
      & \multicolumn{3}{c|}{\textbf{In‑House (1\% Count)}}
      & \multicolumn{3}{c|}{\textbf{UDPET DRF‑100 (1\% Count)}}
      & \multirow{2}{*}{\shortstack{Params\\(M)$\downarrow$}} \\
    \cmidrule(lr){2-4} \cmidrule(lr){5-7} \cmidrule(lr){8-10}
      & SSIM$\uparrow$ & PSNR$\uparrow$ & RMSE$\downarrow$
      & SSIM$\uparrow$ & PSNR$\uparrow$ & RMSE$\downarrow$
      & SSIM$\uparrow$ & PSNR$\uparrow$ & RMSE$\downarrow$
      &  \\
    \midrule
    OSEM            & 0.9078 & 28.35 & 0.0447  & 0.7456 & 23.59 & 0.0745  & 0.7607 & 19.87 & 0.1108  & --     \\
    AutoContextCNN  & 0.9816 & 33.64 & 0.0233  & 0.9339 & 33.66 & 0.0226  & 0.8794 & 26.29 & 0.0541  & 42.56  \\
    DeepPET         & 0.9746 & 30.08 & 0.0331  & 0.8820 & 32.24 & 0.0263  & 0.8218 & 25.28 & 0.0581  & 62.94  \\
    CNNBPnet        & 0.9560 & 30.62 & 0.0329  & 0.9240 & 34.62 & 0.0200  & 0.7750 & 25.06 & 0.0621  & 42.69  \\
    FBPnet          & 0.9327 & 33.62 & 0.0231  & 0.9592 & 34.19 & 0.0210  & 0.8907 & 27.36 & 0.0463  & 21.35  \\
    LCPR‑Net        & 0.9769 & 33.75 & 0.0224  & 0.9222 & 34.95 & 0.0206  & 0.8919 & 27.77 & 0.0446  & 75.93  \\
    Sino‑cGAN       & 0.9641 & 30.76 & 0.0306  & 0.9704 & 33.58 & 0.0223  & 0.8646 & 25.54 & 0.0569  & 46.57  \\
    DGLM\_u         & 0.9785 & 33.58 & 0.0230  & 0.9551 & 32.93 & 0.0245  & 0.8905 & 25.95 & 0.0552  & 0.68   \\
    RED             & 0.9664 & 34.45 & 0.0210  & 0.9472 & 34.15 & 0.0192  & 0.8890 & 26.51 & 0.0474  & 28.93  \\
    \midrule
    \textit{FourierPET} (Ours)
      & \textbf{0.9859} & \textbf{35.36} & \textbf{0.0198}
      & \textbf{0.9740} & \textbf{35.19} & \textbf{0.0188}
      & \textbf{0.9083} & \textbf{27.98} & \textbf{0.0437}
      & \textbf{0.44} \\
    \bottomrule
  \end{tabular}
  \endgroup
\caption{\textbf{Quantitative comparison} of PET reconstruction methods on BrainWeb simulated (20\% Count), In‑House (1\% Count), and UDPET DRF‑100 (1\% Count) datasets.}
\label{tab:recon_comparison_means}
\end{table*}

\begin{figure*}[t]                  
  \centering
  \includegraphics[width=\textwidth]{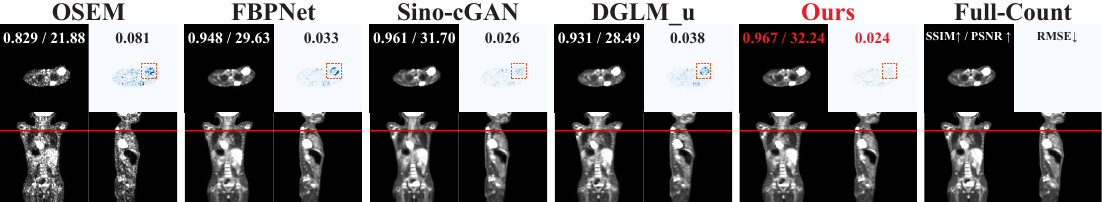}
    \caption{\textbf{Qualitative comparison} on the In‑House dataset. \textbf{Top}: axial slices and corresponding error maps. \textbf{Bottom}: coronal and sagittal views of the same subjects, with red lines indicating axial slice locations. \textbf{Orange rectangles} highlight localized errors in the tumor region of interest (ROI).}
  \label{fig:children_vis}
\end{figure*}

\begin{figure*}[t]                  
  \centering
  \includegraphics[width=0.92\textwidth]{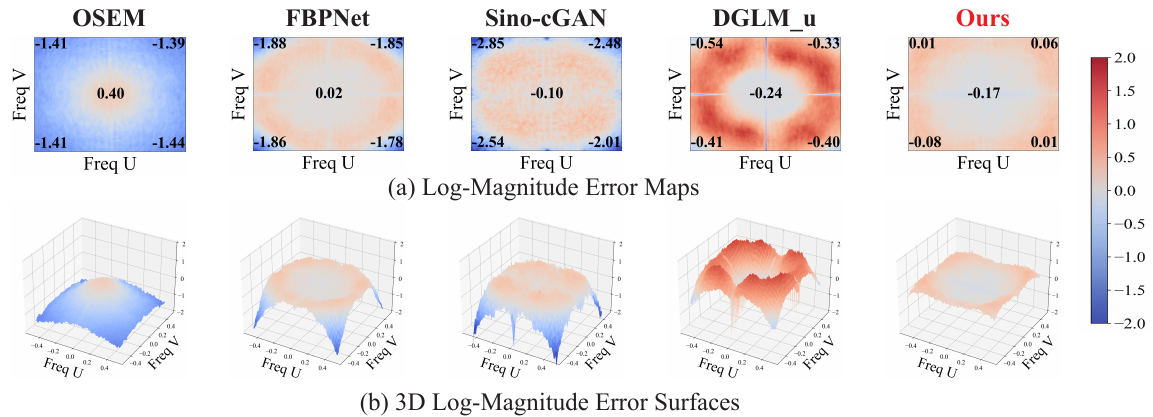}
    \caption{Fourier-domain log-magnitude error analysis on the In-House dataset. \textbf{(a)} 2D error maps and \textbf{(b)} 3D surfaces show the per-frequency deviations between low-count PET reconstructions and the full-count reference in the log-magnitude spectrum. Here, red and blue regions denote overestimation and underestimation in the frequency spectrum, respectively.}
  \label{fig:freq_vis}
\end{figure*}

\subsection{$z$-update: Amplitude-Phase Correction Module}
The $z$‑update step in Eq.~\eqref{eq:sub2} introduces a frequency‑aware regularizer defined as:
\begin{equation}
\label{eq:g}
g(z)=\lambda_a\,\mathcal{R}_{\mathrm{amp}}\bigl(\lvert\mathcal{F}(z)\rvert\bigr)
\;+\;
\lambda_p\,\mathcal{R}_{\mathrm{phase}}\bigl(\angle\mathcal{F}(z)\bigr),
\end{equation}
which targets two characteristic degradation modes in low-count PET: (i) low-frequency amplitude attenuation induced by AC bias, and (ii) high-frequency phase drifts from photon scarcity and Poisson noise. In principle, $z^{k+1}$ should be obtained by solving the proximal mapping:
\begin{equation}
\label{eq:z_prox}
z^{k+1} = \mathrm{prox}_{g/\rho}\bigl(x^{k+1} + u^k\bigr),
\end{equation}
yet the coupled nonlinearity of $g(z)$ precludes a closed-form solution. Iterative solvers can approximate Eq.~\eqref{eq:z_prox}, but they are computationally expensive and fail to explicitly decouple amplitude suppression from phase perturbations. 

To derive a learnable, single-step approximation to Eq.~\eqref{eq:z_prox}, we make two mild assumptions: \textbf{(A1) Band-wise near separability:} after partitioning the frequency spectrum into coarse bands, the penalty in Eq.~\eqref{eq:g} approximately decomposes into band-wise terms acting on localized Fourier spectra. \textbf{(A2) Per-frequency decoupling:} within each band, the proximal operator on the complex spectrum $V=\mathcal{F}(v)$ can be approximated by independent shrinkage on its magnitude and phase components.
Based on these assumptions, we design the Amplitude–Phase Correction Module (APCM), a learnable surrogate that enforces Eq.~\eqref{eq:g} in the spectral domain through three steps (Fig.~\ref{fig:x_z_u}(b)):

\noindent\textbf{(1) Spectral sharding.} 
Given $v = x^{k+1} + u^k$, we first apply a single-level Haar DWT to decompose the image into four sub-bands ${\mathrm{LL}, \mathrm{HL}, \mathrm{LH}, \mathrm{HH}}$. For each band $B$, parallel DWConv layers with dilation rates ${1,2,4}$ extract multi-scale spatial features. A local 2D FFT is then applied to obtain its complex spectrum $V_B = \mathcal{F}(v_B)$, which is further decomposed into amplitude and phase as $(\mathcal{A}_B,\;\Phi_B) = (|V_B|,\; \angle V_B)$. The band-wise structure is preserved throughout, in accordance with assumption (A1).

\noindent\textbf{(2) Directional corrections (Fig.~\ref{fig:PA}).}
Following (A2), amplitude and phase are corrected by separate branches:

\textbf{\emph{Amplitude branch:}}
Each $\mathcal{A}_B$ is processed by a $1{\times}1$ DWConv, followed by batch normalization and GELU activation. To targeted address low-frequency amplitude suppression induced by AC bias, $\mathcal{A}_{\mathrm{LL}}$ is further refined through a two-layer feed-forward network (FFN). A gated residual selectively reinjects the original $\mathcal{A}$, restoring suppressed contrast while avoiding overcorrection, thereby explicitly implementing $\mathcal{R}_{\mathrm{amp}}$ in Eq.~\eqref{eq:g}.

\textbf{\emph{Phase branch:}}
Each phase spectrum $\Phi_B$ is encoded as $(\cos\Phi_B,\sin\Phi_B)$ for numerical stability. 
A high-frequency-focused FFN corrects Poisson-induced angular drifts in $\mathrm{HH}$, followed by cross-band fusion to enforce spectral coherence, thus realizing $\mathcal{R}_{\mathrm{phase}}$.

\noindent\textbf{(3) Spectral fusion.}
Corrected spectra $(\widehat{\mathcal{A}}_B,\widehat{\Phi}_B)$ are combined as
$\widehat{V}_B=\widehat{\mathcal{A}}_B\odot e^{\,\mathrm{i}\widehat{\Phi}_B}$, followed by inverse FFT per band and inverse DWT to reconstruct as
$z^{k+1}=\mathrm{iDWT}\bigl(\{\mathcal{F}^{-1}(\widehat{V}_B)\}_B\bigr)$. This single forward pass efficiently approximates the proximal mapping in Eq.~\eqref{eq:z_prox} under assumptions (A1)–(A2).

\paragraph{Remark.}
APCM serves as a learnable one-step surrogate for $\mathrm{prox}_{g/\rho}$ by (1) partitioning the spectrum into coarse bands where $g$ is approximately separable, and (2) applying per-band amplitude and phase corrections aligned with the penalties in Eq.~\eqref{eq:g}. This design reduces the cost of iterative solvers, preserves physical interpretability, and yields spectrally coherent reconstructions consistent with ADMM.

\subsection{$u$-update: Dual Adjustment Module}

In standard ADMM, the dual variable $u$ accumulates the primal residual $x-z$ with a fixed step size $\mu$:
\begin{equation}
u^{k+1}=u^k+\mu(x^{k+1}-z^{k+1}).
\label{eq:dam}
\end{equation}
However, choosing an appropriate $\mu$ is non-trivial in low-count PET reconstruction:  the primal residual varies considerably across iterations, and a fixed value can either slow down convergence or cause oscillatory behavior. To address this, we introduce the \textbf{Dual Adjustment Module (DAM)} (Fig.~\ref{fig:x_z_u}(c)), which parameterizes $\mu$ as a learnable scalar optimized jointly with other network parameters during unrolled training. This eliminates manual tuning and automatically adapts the dual ascent step size, improving convergence stability without altering the ADMM formulation.

\paragraph{Remark.}
DAM preserves the dual ascent interpretation while enabling automatic, data-driven calibration of the update strength. This removes the need for heuristic tuning and improves convergence stability.
\subsection{Optimization}
To effectively supervise the reconstruction process, we employ a composite loss function between the reconstructed output $x_{\mathrm{out}}$ and the corresponding full-count ground truth $x_{\mathrm{gt}}$, which integrates three complementary components:
\begin{equation}
\mathcal{L}_{\mathrm{total}}
= \lambda_{1}\,\mathcal{L}_{\mathrm{Smooth}\text{-}\mathrm{L1}}
+ \lambda_{2}\,\mathcal{L}_{\mathrm{SSIM}}
+ \lambda_{3}\,\mathcal{L}_{\mathrm{freq}}.
\label{eq:loss}
\end{equation}
where $\mathcal{L}_{\mathrm{Smooth}\text{-}\mathrm{L1}}$ denotes the Smooth L1 loss, $\mathcal{L}_{\mathrm{SSIM}} = 1 - \mathrm{SSIM}(x_{out}, x_{gt})$ denotes the Structural Similarity Index Measure (SSIM) loss~\cite{wang2004image}, and $\mathcal{L}_{\mathrm{freq}} = |\mathcal{F}(x_{out}) - \mathcal{F}(x_{gt})|_1$ is the frequency-domain loss. The weighting coefficients $\lambda_1$, $\lambda_2$, and $\lambda_3$ are empirically set to 0.5, 0.3, and 0.01, respectively, to balance pixel-wise accuracy, structural consistency, and frequency preservation.

\section{Experiment}
\subsection{Experimental Setup}
\paragraph{Datasets.}
We evaluate \textit{FourierPET} on three low-count PET datasets: 
(1) \textbf{BrainWeb}~\cite{aubert2006twenty}: This dataset comprises $20$ simulated brain volumes ($3,200$ slices) with dose levels of $20\%$ and $40\%$. A leave-one-out cross-validation protocol is employed for evaluation.
(2) \textbf{In-house}: This dataset contains $60$ whole-body pediatric PET scans ($40,440$ slices), each paired with synthetically generated acquisitions at $1\%$ and $10\%$ of the standard dose. The data are split into $48$ subjects for training and validation and $12$ subjects for testing.
(3) \textbf{UDPET}~\cite{xue2022cross}: This dataset consists of 206 brain scans ($26{,}368$ slices) acquired with a dose reduction factor (DRF) of 100, of which 170 subjects are used for training and validation, while 36 subjects are reserved for testing.
All experiments use $128\times128$ low-count sinograms as input and full-count OSEM~\cite{hudson1994accelerated} reconstructions as ground truth.
\begin{figure}[t!]      
  \centering
  \includegraphics[width=\columnwidth]{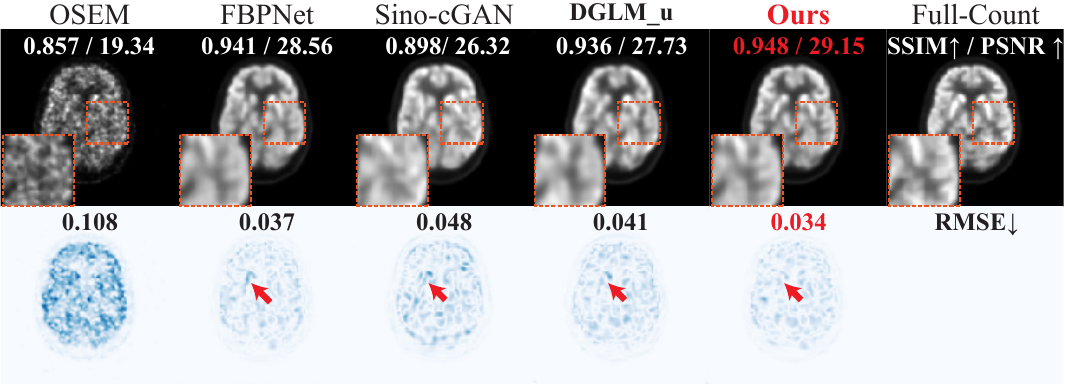}
  \caption{UDPET qualitative results. \textbf{Top}: axial reconstructions with zoomed-in ROIs (orange rectangles); \textbf{Bottom}: error maps computed against the full-count reference.}
  \label{fig:udpet_brain_vis}
\end{figure}
\paragraph{Implementation Details.}
The proposed \textit{FourierPET} architecture is implemented in PyTorch and trained on an NVIDIA RTX 4090 GPU. We adopt the AdamW optimizer with parameters $\beta_{1}=0.9$ and $\beta_{2}=0.999$, and employ a cosine annealing schedule to progressively decrease the learning rate from $1\times10^{-3}$ to $1\times10^{-5}$. To balance reconstruction accuracy and efficiency, we fix the number of unrolled stages at $K=3$ and use a shared internal iteration count $\mathcal{N}=2$ for both SCM and APCM. The reconstruction quality is quantitatively assessed using three widely adopted metrics: PSNR, SSIM, and RMSE.

\subsection{Comparative Evaluation}
We evaluate \textit{FourierPET} against several state-of-the-art PET reconstruction methods, including AutoContextCNN~\cite{xiang2017deep}, DeepPET~\cite{haggstrom2019deeppet}, CNNBPnet~\cite{zhang2020pet}, FBPnet~\cite{wang2020fbp}, LCPR-Net~\cite{xue2021lcpr}, Sino-cGAN~\cite{liu2022deep}, DGLM\_u~\cite{zhang2023deep}, and RED~\cite{ai2025red}. All models are trained on identical data partitions to ensure a fair comparison, while their original loss functions and architectural settings are preserved. As summarized in \textbf{Tab.~\ref{tab:recon_comparison_means}}, \textit{FourierPET} consistently achieves the highest PSNR and SSIM, along with the lowest RMSE, while requiring fewer trainable parameters than competing methods.

Qualitative comparisons are provided in \textbf{Fig.~\ref{fig:children_vis}} and \textbf{Fig.~\ref{fig:udpet_brain_vis}}, showcasing reconstructed images and corresponding error maps. Even under extreme low-count conditions, \textit{FourierPET} preserves structure and contrast with minimal artifacts. Furthermore, \textbf{Fig.~\ref{fig:freq_vis}} visualizes log-magnitude errors in the Fourier domain. Across both 2D and 3D visualizations, \textit{FourierPET} demonstrates the lowest spectral distortion, highlighting its superior frequency-domain fidelity. 

\subsection{Ablation Studies and Analyses}
\paragraph{Effectiveness of Core Components.}
We begin with a baseline unrolled ADMM network in which both the $x$-update and $z$-update steps are implemented using three-layer $3{\times}3$ convolutional blocks with LeakyReLU activations. To assess the individual contributions of our proposed modules, we successively replace the $x$-update with SCM and the $z$-update with APCM. As summarized in
\textbf{Tab.~\ref{tab:ablation1}}, each component yields a noticeable performance improvement over the baseline, and their combination restores the full advantage of \textit{FourierPET}, highlighting their complementary effects.

\begin{table}[t]
  \centering
  \scriptsize
  \setlength{\tabcolsep}{2.6pt}
  \begin{tabular}{ccccc|ccc}
    \toprule
    \multirow{2}{*}{\textbf{SCM}} & 
    \multirow{2}{*}{\textbf{APCM}} &
    \multicolumn{3}{c|}{\textbf{In‑House}} &
    \multicolumn{3}{c}{\textbf{UDPET}} \\
    \cmidrule(lr){3-5}\cmidrule(lr){6-8}
    & & SSIM$\uparrow$ & PSNR$\uparrow$ & RMSE$\downarrow$
      & SSIM$\uparrow$ & PSNR$\uparrow$ & RMSE$\downarrow$ \\
    \midrule
              &               & 0.940 & 33.15 & 0.0237 & 0.891 & 25.95 & 0.055 \\
    \checkmark &               & 0.971 & 34.62 & 0.0200 & 0.894 & 27.36 & 0.046 \\
              & \checkmark     & 0.967 & 34.05 & 0.0210 & 0.880 & 26.22 & 0.053 \\
    \checkmark & \checkmark     & \textbf{0.974} & \textbf{35.19} & \textbf{0.0190} & \textbf{0.908} & \textbf{27.98} & \textbf{0.044} \\
    \bottomrule
  \end{tabular}
    \caption{Ablation study of our core components.}
  \label{tab:ablation1}
\end{table}
\begin{table}[t]
  \centering
  \small
  \setlength{\tabcolsep}{5.5pt}
  \begin{tabular}{lccc}
    \toprule
    \textbf{SCM Variant} & \textbf{SSIM $\uparrow$} & \textbf{PSNR $\uparrow$} & \textbf{RMSE $\downarrow$} \\
    \midrule
    w/o $\mathrm{A}\!^\top$       & 0.8328 & 22.55 & 0.0849 \\
    w/o SSFNO                     & 0.9530 & 33.69 & 0.0224 \\
    w/o Spatial Module            & 0.9681 & 34.43 & 0.0205 \\
    Full SCM                      & \textbf{0.9740} & \textbf{35.19} & \textbf{0.0188} \\
    \bottomrule
  \end{tabular}
    \caption{Ablation study of SCM components.}
  \label{tab:scm_ablation}
\end{table}

\begin{table}[t!]
  \centering
  \scriptsize
  \setlength{\tabcolsep}{2.6pt}
  \begin{tabular}{ccccc|ccc}
    \toprule
    \multirow{2}{*}{\textbf{Phase}} & 
    \multirow{2}{*}{\textbf{Amp}} &
    \multicolumn{3}{c|}{\textbf{In‑House}} & 
    \multicolumn{3}{c}{\textbf{UDPET}} \\
    \cmidrule(lr){3-5} \cmidrule(lr){6-8}
    & & SSIM$\uparrow$ & PSNR$\uparrow$ & RMSE$\downarrow$
      & SSIM$\uparrow$ & PSNR$\uparrow$ & RMSE$\downarrow$ \\
    \midrule
    \checkmark &             & \textbf{0.968} & 33.95 & 0.0216 
                         & \textbf{0.886} & 26.01 & 0.054 \\
              & \checkmark  & 0.958 & 34.01 & 0.0212 
                         & 0.878 & 26.00 & 0.054 \\
    \checkmark & \checkmark & 0.967 & \textbf{34.05} & \textbf{0.0210} 
                         & 0.880 & \textbf{26.22} & \textbf{0.053} \\
    \bottomrule
  \end{tabular}
    \caption{Effect of phase (\textbf{Phase}) and amplitude (\textbf{Amp}) branches in APCM.}
  \label{tab:apcm_ablation}
\end{table}

\begin{figure}[t]      
  \centering
  \includegraphics[width=\columnwidth]{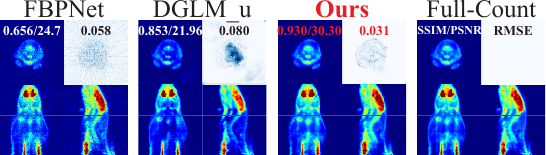}
    \caption{{Zero‑shot adaptation} on low‑count, \emph{in vivo} mouse PET. \textbf{Top}: axial reconstructions and corresponding error maps. \textbf{Bottom}: coronal and sagittal views of the same subject. Reconstructions are generated from a model pretrained on human data, shown alongside full-count references.}
  \label{fig:zero_shot}
\end{figure}

\paragraph{Efficacy of SCM Submodules.}
We conduct ablation experiments on the in-house dataset to assess the contributions of the SSFNO, the spatial module, and the $\mathrm{A}^{\top}$-based constraint within the SCM. As reported in \textbf{Tab.~\ref{tab:scm_ablation}}, removing any of these submodules leads to a noticeable degradation in reconstruction quality, indicating that all three are essential for preserving spectral coherence, maintaining structural fidelity, and ensuring measurement consistency.

\paragraph{Effect of Phase/Amplitude Branches in APCM.}
We perform an ablation study on the APCM by isolating its phase and amplitude branches to evaluate their individual contributions. As shown in \textbf{Tab.~\ref{tab:apcm_ablation}}, the phase branch primarily enhances structural fidelity by stabilizing high-frequency components, leading to higher SSIM values. The amplitude branch focuses on low-frequency corrections, effectively reducing global bias and improving both PSNR and RMSE. Combining both branches yields the best overall performance, highlighting the importance of joint spectral targeting. The slight reduction in SSIM observed in the combined setup suggests a trade-off, where amplitude correction enhances global fidelity at the expense of a marginal loss in local structural sharpness.

\paragraph{Analysis of Zero-Shot Generalization.}
To evaluate the model’s generalization capability, we perform a zero-shot adaptation experiment by directly applying models pretrained on human PET data to \emph{in vivo} mouse scans acquired under a low-count protocol. As illustrated in \textbf{Fig.~\ref{fig:zero_shot}}, \textit{FourierPET} maintains high reconstruction quality across domains, highlighting its robustness and strong potential for cross-species transfer and future clinical translation.


\section{Conclusion}
This paper presents a novel frequency-domain perspective for low-count PET reconstruction, establishing a direct link between data degradations and their spectral signatures: high-frequency phase drift induced by Poisson noise and photon scarcity, and low-frequency amplitude suppression caused by AC bias. Building on this observation, we propose \textit{FourierPET}, an ADMM-unrolled framework that exploits spectral decomposition to perform targeted frequency-domain corrections. The proposed SCM, APCM, and DAM modules collaboratively enforce data fidelity, selectively rectify spectral distortions, and guarantee convergence stability. Extensive experiments on diverse datasets validate the effectiveness, robustness, and generalizability of \textit{FourierPET}, highlighting its potential for efficient and high-quality low-count PET reconstruction.

\section*{Appendix}

\begin{itemize}
  \item \textbf{Appendix A: Related Work}

  \item \textbf{Appendix B: Additional Experiments}
  \begin{itemize}
    \item B.1 \quad Performance under Higher Count Levels  
    \item B.2 \quad Supplementary Visual Comparisons
  \end{itemize}

  \item \textbf{Appendix C: Additional Analyses}
  \begin{itemize}
    \item C.1 \quad Effect of Frequency--Selective Correction 
    \item C.2 \quad Effect of DAM on Convergence 
    \item C.3 \quad Impact of Unrolled Stages $K$ 
    \item C.4 \quad Impact of Module Depth $\mathcal{N}$  
    \item C.5 \quad Loss Formulations Analysis
  \end{itemize}

  \item \textbf{Appendix D: Limitations}
\end{itemize}

\section*{Appendix A: Related Work}

\subsection{Low‐count PET Reconstruction}
Deep learning has significantly advanced low-count PET reconstruction by addressing key limitations of traditional algorithms such as MLEM~\cite{shepp2007maximum} and OSEM~\cite{hudson1994accelerated}, which are constrained by slow convergence and poor noise suppression. Existing learning-based methods fall into two main categories.

The first involves post-reconstruction enhancement, where networks refine images reconstructed by conventional algorithms (e.g. OSEM) to suppress noise and restore contrast~\cite{gao2025multistage,cui20243d,TangHQ25,zeng2025multi,cui2024mcad}. While simple to deploy, these methods are fundamentally limited by upstream reconstruction quality.

The second category comprises end-to-end approaches that learn direct mappings from measured sinograms to reconstructed images. Within this group, direct regression models~\cite{cui2024prior,jiang2024end,webber2025likelihood,shen2024imagpose,TangYLT22,0007LYYL023,ai2025red} predict images in a one-shot manner, whereas unrolled optimization frameworks~\cite{zhang2023deep,xie2025prompt,hu2022transem} mimic iterative solvers with learnable update rules, allowing stronger integration of imaging physics. 

Despite their success, most existing models rely on data-driven image regression, which often lacks physical interpretability and provides limited insight into the underlying degradation mechanisms, thereby constraining their robustness and transparency. In contrast, our approach adopts a frequency-domain perspective that explicitly separates low-count degradations into distinct spectral components, enabling interpretable, targeted correction and improved robustness.

\subsection{Spectral‑Domain Image Restoration}
Spectral-based techniques have achieved remarkable success in both natural image restoration~\cite{zhou2023fourmer,yu2022deep,zhou2024seeing,TangLPT20,he2025unfoldir,feijoo2025darkir,cui2025eenet,gao2024efficient,shen2025imagdressing,shenlong,li2024fouriermamba} and medical imaging, largely by decoupling degradations in the spectral domain. In MRI, phase unwrapping decouples field inhomogeneity from magnitude contrast~\cite{zhu2025dip,haacke1999magnetic,shen2025imaggarment,haller2021susceptibility}, and dual-energy CT leverages spectral attenuation differences to correct for beam hardening~\cite{tatsugami2022dual,Alvarez1976}. While such strategies are well established in these domains, their adoption in PET remains limited. However, the fundamentally different physics of low-count PET introduces degradation patterns that are more complex and entangled, limiting the direct applicability of existing methods.
Inspired by prior work, we find that these degradations nonetheless exhibit a separation in the Fourier domain. Leveraging the insight, we propose \textit{FourierPET}, a frequency-aware reconstruction framework that applies directional priors within targeted spectral bands while preserving global consistency.

\section*{Appendix B: Additional Experiments}
\subsection{B.1 Performance under Higher Count Levels}
\begin{table*}[t]
  \centering
  \begingroup
  \setlength{\tabcolsep}{8pt}
  \renewcommand{\arraystretch}{1.1}
  \setlength{\extrarowheight}{0pt}
  \footnotesize
  \begin{tabular}{l|ccc|ccc}
    \toprule
    \textbf{Method}
      & \multicolumn{3}{c|}{\textbf{BrainWeb (40\% Count)}}
      & \multicolumn{3}{c}{\textbf{In‑House (10\% Count)}} \\
    \cmidrule(lr){2-4} \cmidrule(lr){5-7}
      & SSIM$\uparrow$ & PSNR$\uparrow$ & RMSE$\downarrow$
      & SSIM$\uparrow$ & PSNR$\uparrow$ & RMSE$\downarrow$ \\
    \midrule
    OSEM            & 0.9267 & 29.34 & 0.0398  & 0.9082 & 31.36 & 0.0304  \\
    autoContextCNN  & 0.9834 & 37.15 & 0.0164  & 0.9730 & 38.15 & 0.0135  \\
    DeepPET         & 0.9742 & 29.78 & 0.0341  & 0.9409 & 34.38 & 0.0207  \\
    CNNBPnet        & 0.9524 & 33.24 & 0.0234  & 0.8060 & 34.77 & 0.0199  \\
    FBPnet          & 0.9683 & 35.58 & 0.0187  & 0.9832 & 39.30 & 0.0116  \\
    LCPR--Net        & 0.9759 & 33.63 & 0.0230  & 0.9650 & 37.98 & 0.0144  \\
    Sino--CGAN       & 0.9577 & 30.65 & 0.0313  & 0.9534 & 36.28 & 0.0191  \\
    DGLM\_u         & 0.9811 & 34.33 & 0.0210  & 0.9762 & 36.78 & 0.0157  \\
    RED             & 0.9599 & 36.75 & 0.0180  & 0.9501 & 37.33 & 0.0149  \\
    \textit{FourierPET} & \textbf{0.9918} & \textbf{37.38} & \textbf{0.0158}  & \textbf{0.9852} & \textbf{40.45} & \textbf{0.0105}  \\
    \bottomrule
  \end{tabular}
  \endgroup
\caption{\textbf{Quantitative comparison} of PET reconstruction methods on BrainWeb simulated (40\% Count) and In‑House (10\% Count) datasets.}
\label{tab:recon_comparison_40k10p}
\end{table*}
To assess robustness across varying noise regimes, we evaluate all methods on higher-count acquisitions: BrainWeb at 40\% and In-House at 10\% count levels, both reflecting clinically favorable protocols with improved signal fidelity.

As summarized in \textbf{Tab.~\ref{tab:recon_comparison_40k10p}}, \textit{FourierPET} maintains state-of-the-art performance across both datasets. It achieves a PSNR of 37.38 and an SSIM of 0.9918 on BrainWeb, and reaches 40.45 PSNR with 0.0105 RMSE on In-House, consistently outperforming all baselines. These results indicate that \textit{FourierPET} not only addresses extreme low-count degradation but also scales effectively to higher-quality inputs without saturation or performance collapse.

\subsection{B.2 Supplementary Visual Comparisons}
\begin{figure*}[t]                  
  \centering
  \includegraphics[width=2\columnwidth]{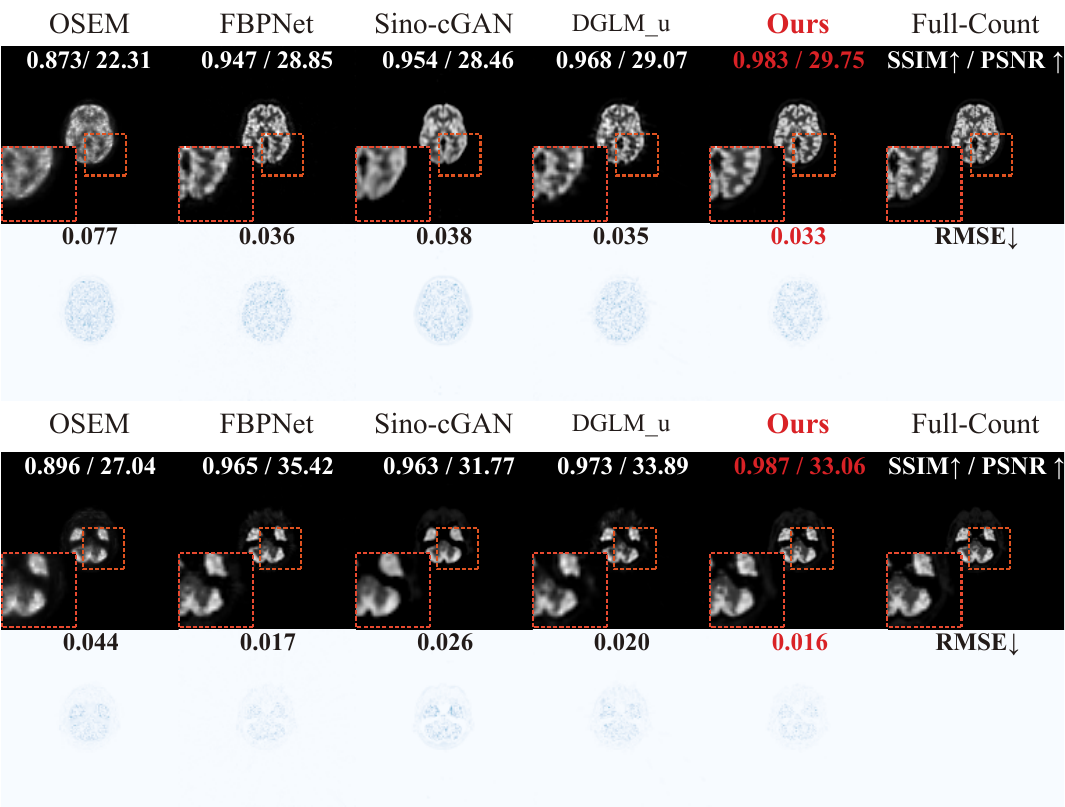}\vspace{-1mm}
    \caption{\textbf{Supplementary qualitative results} on the BrainWeb dataset (20\% count). \textbf{Top}: axial reconstructions with zoomed-in ROIs (orange boxes). \textbf{Bottom}: corresponding error maps computed against the full-count reference.}
  \label{fig:vis_appendix_brainweb}\vspace{-2mm}
\end{figure*}

\begin{figure*}[t]                  
  \centering
  \includegraphics[width=1\textwidth]{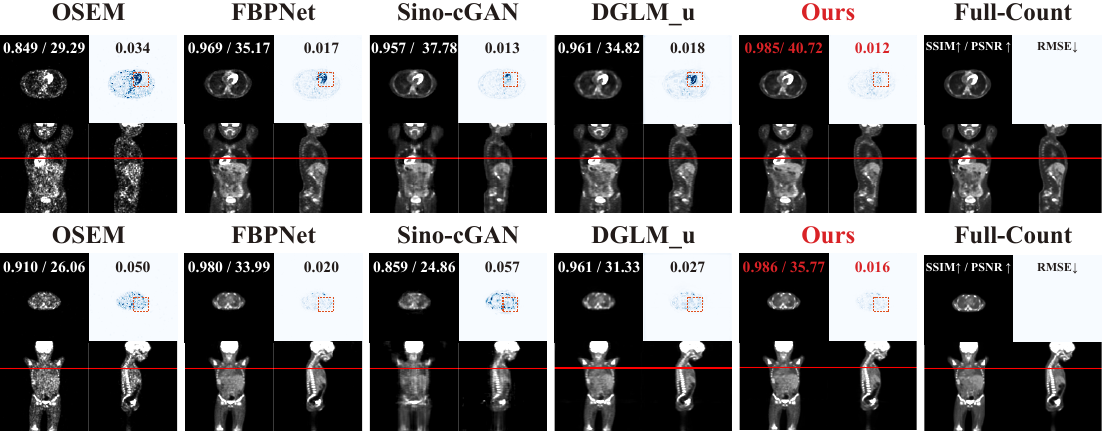}\vspace{-1mm}
    \caption{\textbf{Supplementary qualitative comparison} on the In‑House dataset (1\% count). \textbf{Top}: axial slices and corresponding error maps. \textbf{Bottom}: coronal and sagittal views of the same subjects, with red lines indicating axial slice locations. \textbf{Orange rectangles} highlight localized errors in the tumor region of interest (ROI).}
  \label{fig:vis_appendix_children1}\vspace{-2mm}
\end{figure*}

\begin{figure*}[t]                  
  \centering
  \includegraphics[width=1.5\columnwidth]{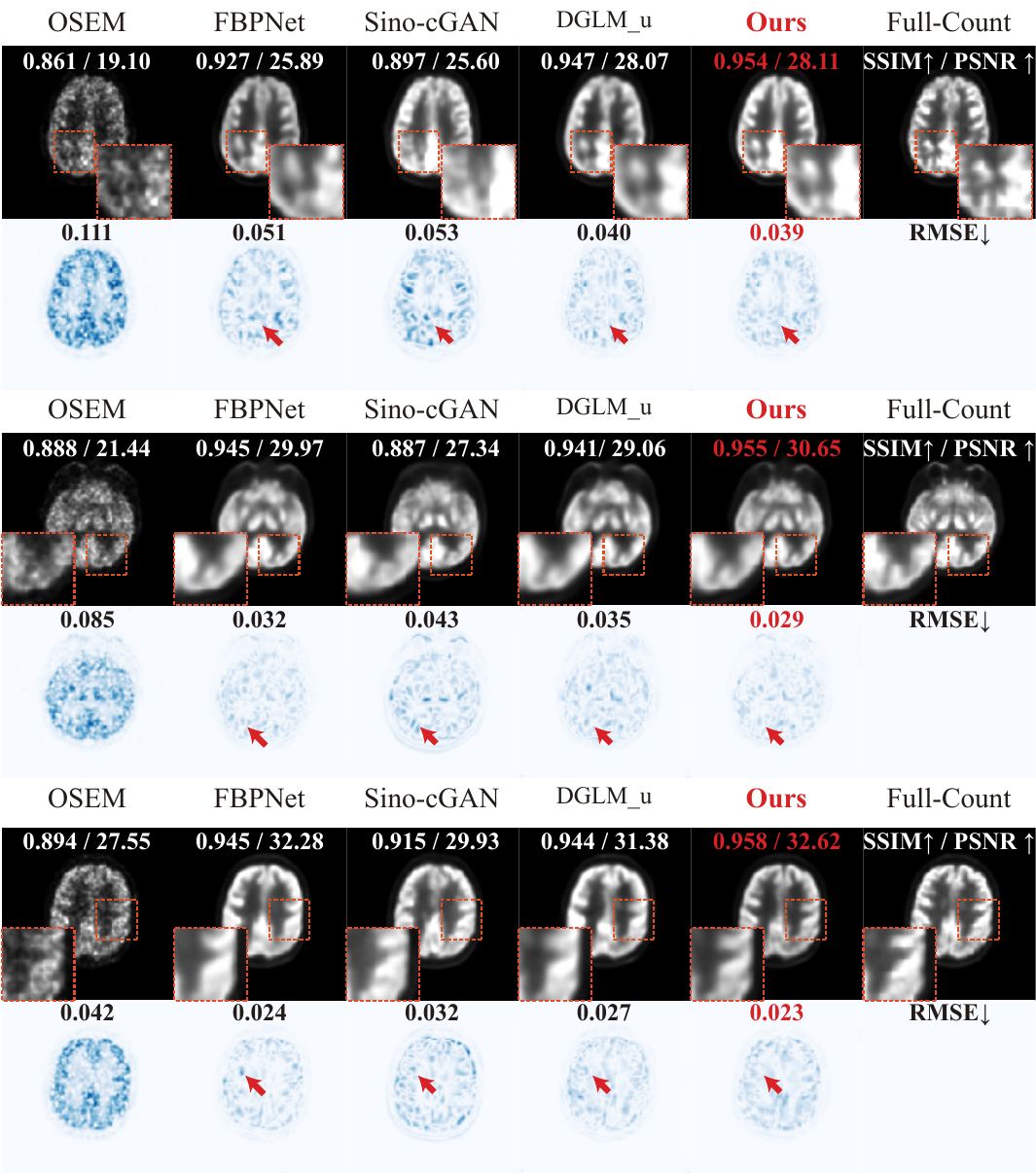}\vspace{-1mm}
    \caption{\textbf{Supplementary qualitative comparison} on the UDPET dataset (1\% count). \textbf{Top}: axial reconstructions with zoomed-in ROIs (orange boxes). \textbf{Bottom}: corresponding error maps computed against the full-count reference.}
  \label{fig:vis_appendix_udpet}\vspace{-2mm}
\end{figure*}

\begin{figure*}[t]                  
  \centering
  \includegraphics[width=0.8\textwidth]{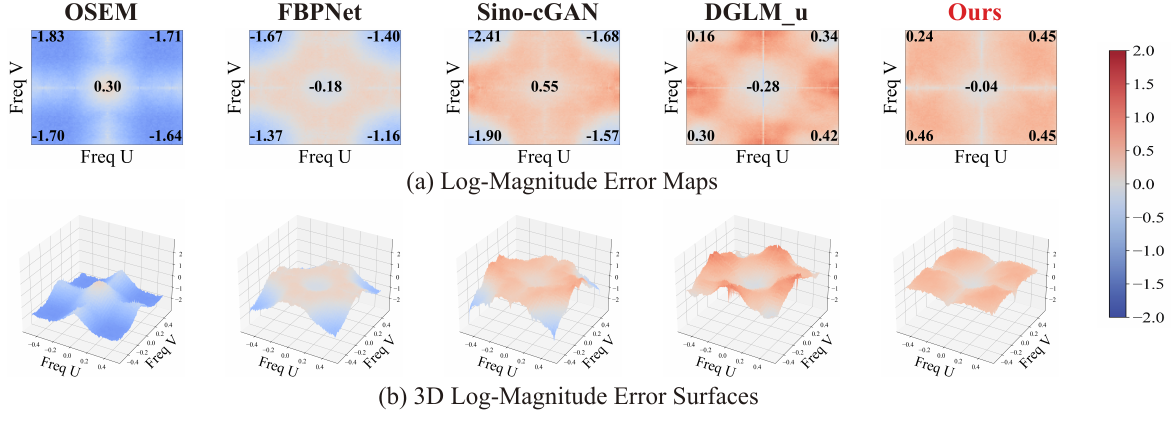}\vspace{-1mm}
    \caption{\textbf{Supplementary Fourier-domain log-magnitude error analysis} on the UDPET dataset (1\% count). \textbf{(a)} 2D error maps and \textbf{(b)} 3D surfaces show the frequency-wise deviations between low-count PET reconstructions and the full-count reference in the log-magnitude spectrum. Red and blue regions denote overestimation and underestimation in the frequency spectrum, respectively.}
  \label{fig:vis_appendix_udpet_freq}\vspace{-2mm}
\end{figure*}

We provide additional qualitative comparisons in \textbf{Fig.~\ref{fig:vis_appendix_brainweb}}, \textbf{Fig.~\ref{fig:vis_appendix_children1}}, and \textbf{Fig.~\ref{fig:vis_appendix_udpet}}, corresponding to the BrainWeb (20\% count), In-House (1\%), and UDPET (1\%) datasets, respectively. Each figure displays reconstructed slices alongside full-count references and error maps, enabling visual evaluation of structural fidelity and noise suppression. Additionally, \textbf{Fig.~\ref{fig:vis_appendix_udpet_freq}} presents 2D and 3D log-magnitude error visualizations in the Fourier domain on the UDPET dataset.

Across all datasets, \textit{FourierPET} consistently preserves structural detail and radiotracer distribution under extreme low-count conditions. Compared to competitive methods, it better maintains lesion sharpness, suppresses noise without over-smoothing, and visually aligns more closely with the ground truth reference, highlighting its robustness in both spatial and frequency domains.

\section*{Appendix C: Additional Analyses}
\subsection{C.1 Effect of Frequency-Selective Correction}
\paragraph{Targeted vs. Full-Band Frequency Correction in APCM.}
Although spectral degradation spans the entire frequency domain, our APCM deliberately restricts correction to low-frequency amplitude and high-frequency phase components, guided by the empirical patterns in \textbf{Fig.~\ref{fig:total_experiment}}. This raises the question of whether uniform correction across all frequency bands would yield better results.

\begin{figure}[t]
  \centering
  \includegraphics[width=\columnwidth]{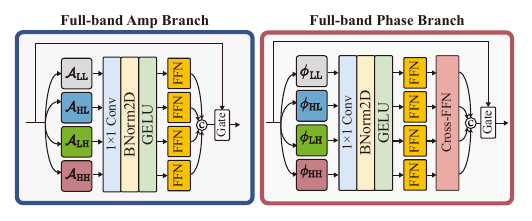}\vspace{-1mm}
  \caption{\textbf{Full-band APCM variants.} Left: amplitude branch applies uniform correction to all sub-bands. Right: phase branch processes all phase components equally. Our design instead targets $\mathcal{A}_{LL}$ and $\mathcal{\phi}_{HH}$.}
  \label{fig:Full_PA}\vspace{-2mm}
\end{figure}

To investigate this, we compare our targeted design (\textbf{Fig.~\ref{fig:PA}}) with a full-band variant that applies identical correction to all sub-bands, thereby discarding frequency selectivity (\textbf{Fig.~\ref{fig:Full_PA}}). As shown in \textbf{Tab.~\ref{tab:target_vs_full}}, the full-band variant achieves slightly lower RMSE ($0.0187$ vs. $0.0190$), reflecting improved global intensity fit. However, it exhibits reduced SSIM ($0.9709$ vs. $0.9740$), indicating diminished structural fidelity. This trade-off suggests that indiscriminate correction may overwrite clean spectral regions—particularly mid-frequency components—thereby weakening directional priors.

Our targeted approach aligns with the spectral deviation profiles (Fig.~\ref{fig:total_experiment}(c)) and underlying assumptions (A1–A2): by focusing on frequency bands most affected by low-count degradation, where low-frequency amplitude suffers from AC bias and high-frequency phase from Poisson noise, our method avoids unnecessary correction of stable components. This selective strategy preserves anatomical structure and encourages more efficient, localized learning. 

\begin{table}[t]
  \centering
  \begin{tabular}{lccc}
    \toprule
    \textbf{Method} & \textbf{SSIM}$\uparrow$ & \textbf{PSNR}$\uparrow$ & \textbf{RMSE}$\downarrow$ \\
    \midrule
    Full-band Correction        & 0.9709 & \textbf{35.22} & \textbf{0.0187} \\
    Targeted Correction (Ours)  & \textbf{0.9740} & 35.19 & 0.0190 \\
    \bottomrule
  \end{tabular}
\caption{\textbf{Targeted vs. full-band correction in APCM.} Targeted correction yields higher SSIM with comparable RMSE, demonstrating the advantage of focusing on degraded spectral regions.}
  \label{tab:target_vs_full}
\end{table}

\subsection{C.2 Effect of DAM on Convergence}
\begin{figure}[ht!]
  \centering
    \includegraphics[width=\linewidth]{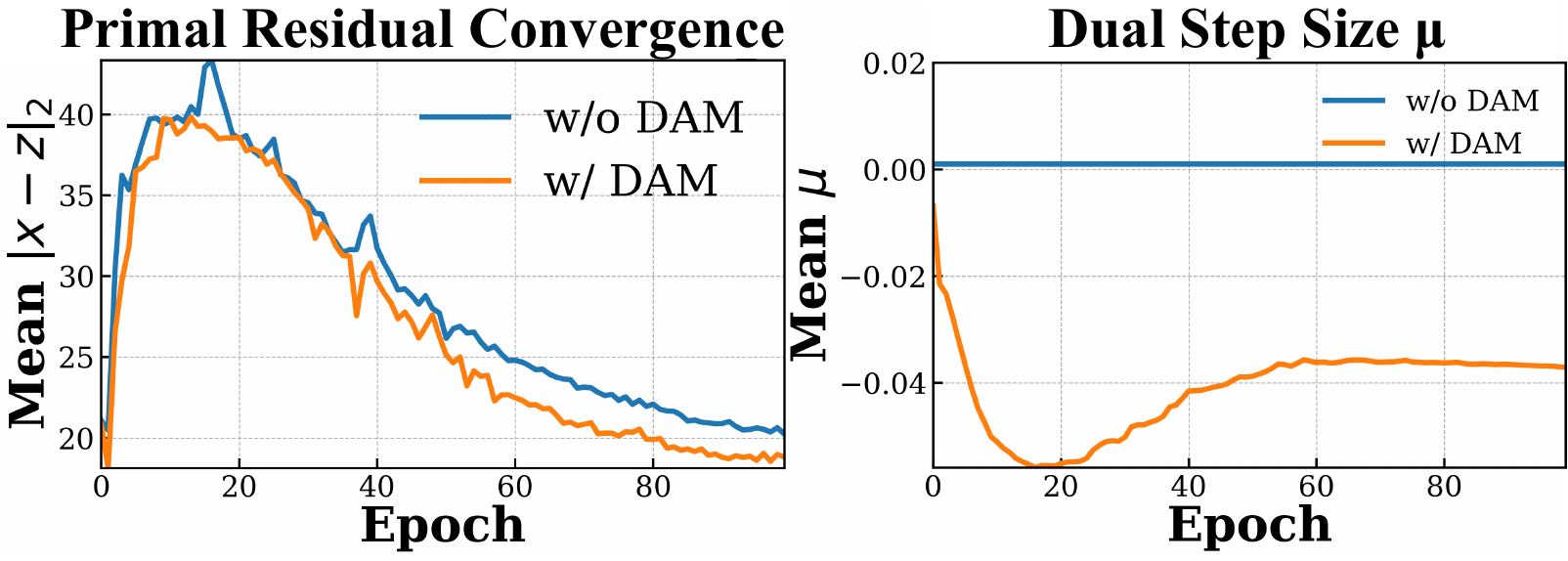}
    \caption{
    \textbf{Effect of DAM on convergence.}
    \textbf{Left:} Primal residual $\|x - z\|_2$ over training epochs. DAM accelerates convergence and stabilizes updates.
    \textbf{Right:} Learned step size $\mu$ adapts dynamically, unlike the fixed $\mu$ in vanilla ADMM.
    }\vspace{-2mm}
  \label{fig:dam_convergence}
\end{figure}
To evaluate the impact of our proposed Dual Adjustment Module (DAM), we build upon the ADMM formulation in Eq.~\eqref{eq:subs}, which decomposes the reconstruction problem in Eq.~\eqref{eq:objective} into alternating updates over $x$, $z$, and the dual variable $u$. While effective, conventional ADMM employs a fixed step size $\mu$ (with~$\mu = 1/\rho$), which often leads to slow or unstable convergence in ill-posed settings like low-count PET reconstruction.

DAM mitigates this limitation by learning an adaptive step size $\mu$ to dynamically modulate the dual update in Eq.~\eqref{eq:dam}. We evaluate its effectiveness by monitoring the primal residual $\|x - z\|_2$, which serves as a standard ADMM convergence metric that reflects the consensus constraint $x = z$, as well as the evolution of $\mu$ during training.

As shown in Fig.~\ref{fig:dam_convergence}, DAM accelerates and stabilizes the decrease of the primal residual, suggesting improved consistency between the primal and auxiliary variables across iterations. The learned step size $\mu$ adjusts rapidly during the early training stages and gradually converges to a stable value, allowing \textit{FourierPET} to adaptively balance speed and stability in optimization. This data-driven adjustment enhances the robustness of optimization and eliminates the need for manual step size tuning.

\subsection{C.3 Impact of Unrolled Stages $K$}
We evaluate how the number of unrolled stages $K$ affects reconstruction performance and inference cost in \textit{FourierPET}. As shown in \textbf{Tab.~\ref{tab:unrolling_ablation}}, increasing $K$ yields consistent gains in SSIM and PSNR, but also leads to higher parameter count and longer inference time. Here, \textit{Time} denotes the average duration required to reconstruct a whole-body PET scan comprising 674 axial slices. All timings are measured on a single NVIDIA RTX 4090 GPU.
\renewcommand{\arraystretch}{0.9} 
\begin{table}[!t]
  \centering
  \setlength{\tabcolsep}{5pt}
  \begin{tabular}{c|ccc|c|c}
    \toprule
    \textbf{$K$} 
    & SSIM$\uparrow$ & PSNR$\uparrow$ & RMSE$\downarrow$ 
    & Param (M) & Time (s) \\
    \midrule
    3  & 0.9740 & 35.19 & 0.0188 & \textbf{0.44} & \textbf{20.25} \\
    5  & 0.9754 & 35.32 & 0.0184 & 0.73 & 34.58 \\
    10 & \textbf{0.9790} & \textbf{36.01} & \textbf{0.0170} & 1.46 & 58.87 \\
    \bottomrule
  \end{tabular}
    \caption{Effect of unrolling stages $K$ on reconstruction performance, parameter count, and average inference time per whole-body scan (674 slices, In‑House dataset).}
  \label{tab:unrolling_ablation}
\end{table}
\subsection{C.4 Impact of Module Depth $\mathcal{N}$}
As shown in \textbf{Fig.~\ref{fig:x_z_u}}, we define $\mathcal{N}$ as the number of stacked spectral blocks per iteration, which determines the depth of both the SSFNO module in the $x$-update (SCM) and the dual-branch module in the $z$-update (APCM). Each \textit{FourierPET} iteration applies $\mathcal{N}$ SSFNO units to enforce spectral consistency and $\mathcal{N}$ amplitude and phase correction blocks for targeted frequency-domain modulation.

We evaluate $\mathcal{N} \in \{2, 4, 6\}$ and report results in \textbf{Tab.~\ref{tab:ssfno_apcm_ablation}}. Increasing $\mathcal{N}$ improves SSIM and PSNR up to $\mathcal{N}=4$, beyond which accuracy saturates while parameter count and reconstruction time increase. In practice, we adopt $\mathcal{N}=2$ to minimize computation without significant performance loss.
\renewcommand{\arraystretch}{0.9} 
\begin{table}[!t]
  \centering
  \setlength{\tabcolsep}{5pt}
  \begin{tabular}{c|ccc|c|c}
    \toprule
    \textbf{$\mathcal{N}$} 
    & SSIM$\uparrow$ & PSNR$\uparrow$ & RMSE$\downarrow$
    & Param (M) & Time (s) \\
    \midrule
    2 & 0.9740 & 35.19   & 0.0188  & \textbf{0.44} & \textbf{20.25} \\
    4 & \textbf{0.9767} & \textbf{35.73}   & \textbf{0.0176}  & 0.88  & 32.82 \\
    6 & 0.9765 & 35.70   & 0.0177  & 1.28  & 49.61 \\
    \bottomrule
  \end{tabular}
  \caption{Impact of module depth $\mathcal{N}$ on reconstruction quality and computational cost. We report SSIM, PSNR, parameter count, and reconstruction time per whole-body PET scan on the In-House dataset. $\mathcal{N}$ controls the number of SSFNO blocks in SCM and amplitude/phase blocks in APCM.}
  \label{tab:ssfno_apcm_ablation}
\end{table}

\subsection{C.5 Impact of Loss Components}
We ablate the composite loss $\mathcal{L}_{\text{total}}$ in Eq.~\eqref{eq:loss} to assess the contribution of each term. As shown in \textbf{Tab.~\ref{tab:loss_ablation}}, the full combination of $\mathcal{L}_{\mathrm{Smooth}\text{-}L1}$, $\mathcal{L}_{\mathrm{SSIM}}$, and $\mathcal{L}_{\mathrm{freq}}$ yields the best reconstruction performance. Excluding $\mathcal{L}_{\text{freq}}$ noticeably degrades spectral fidelity, highlighting its role in suppressing aliasing and preserving high-frequency structural details.

\begin{table}[t]
  \centering
  \small
  \setlength{\tabcolsep}{4.5pt}
  \begin{tabular}{ccc|ccc}
    \toprule
    $\mathcal{L}_{\mathrm{Smooth}\text{-}L1}$ & $\mathcal{L}_{\mathrm{SSIM}}$ & $\mathcal{L}_{\mathrm{freq}}$
    & SSIM$\uparrow$ & PSNR$\uparrow$ & RMSE$\downarrow$ \\
    \midrule
    \checkmark &             &             & 0.9576 & 34.64 & 0.0200 \\
    \checkmark &             & \checkmark  & 0.9540 & 34.67 & 0.0199 \\
    \checkmark & \checkmark  &             & 0.9737 & 34.92 & 0.0191 \\
    \checkmark & \checkmark  & \checkmark  & \textbf{0.9740} & \textbf{35.19} & \textbf{0.0188} \\
    \bottomrule
  \end{tabular}
  \caption{Ablation study of the loss components in Eq.~\eqref{eq:loss}. Results are reported on the In‑House dataset.}
  \label{tab:loss_ablation}
\end{table}

\section*{Appendix D: Limitations}
Despite the promising performance of \textit{FourierPET}, several limitations merit discussion. 

First, the APCM serves as a learnable surrogate to the proximal operator of the compound spectral regularizer $g(z)$ in Eq.~\eqref{eq:z_prox}, which lacks a closed-form due to its nonlinear coupling. While APCM assumes band-wise separability and amplitude–-phase decoupling, these assumptions currently lack formal justification. To improve its interpretability and reproducibility, we plan to derive APCM from a principled variational formulation and quantify the approximation gap via error-bound analysis.

Second, current evaluations rely on pixel-level metrics and SUV error, which fail to reflect clinical relevance such as lesion detectability or diagnostic confidence. We aim to incorporate radiologist studies and downstream diagnostic tasks to strengthen clinical significance.

Third, current \textit{FourierPET} depends on fixed spectral transforms, assuming consistent frequency characteristics across imaging protocols and anatomies. We will explore adaptive or learnable spectral decompositions to improve robustness under anatomical and tracer variability.


\clearpage

\clearpage

\bibliography{reference.bib}

\end{document}